\documentclass[twocolumn]{aastex63}

\newcommand{\msun}{${\rm M_{\sun}}$}

\def\ltsima{$\; \buildrel < \over \sim \;$}
\def\simlt{\lower.5ex\hbox{\ltsima}}
\def\gtsima{$\; \buildrel > \over \sim \;$}
\def\simgt{\lower.5ex\hbox{\gtsima}}
\def\kms{{\rm\,km\,s^{-1}}}

\def\masyr{{\rm\,mas/yr}}
\def\kpc{{\rm\,kpc}}

\def\msun{{\rm\,M_\odot}}

\def\g{{\rm\,g}}

\def \r {r$^{1/4}$ }


\def\deg{^\circ}
\def\degg{\hbox{$\null^\circ$\hskip-3pt .}}

\def\s{\ifmmode \widetilde \else \~\fi}
\def\={\overline}

\def\spose#1{\hbox to 0pt{#1\hss}}

\def\lta{\mathrel{\spose{\lower 3pt\hbox{$\mathchar"218$}}
     \raise 2.0pt\hbox{$\mathchar"13C$}}}
\def\gta{\mathrel{\spose{\lower 3pt\hbox{$\mathchar"218$}}
     \raise 2.0pt\hbox{$\mathchar"13E$}}}
\def\Dt{\spose{\raise 1.5ex\hbox{\hskip3pt$\mathchar"201$}}}    
\def\dt{\spose{\raise 1.0ex\hbox{\hskip2pt$\mathchar"201$}}}    

\def\r{${\rm r^{1/4}}$}

\def\dotsfill{\leaders\hbox to 1em{\hss.\hss}\hfill}
\def\sun{\odot}
\def\earth{\oplus}

\def\Gyr{{\rm\,Gyr}}

\def\ltsima{$\; \buildrel < \over \sim \;$}
\def\gtsima{$\; \buildrel > \over \sim \;$}
\def\lsim{\lower.5ex\hbox{\ltsima}}
\def\gsim{\lower.5ex\hbox{\gtsima}}
\def\lapp{\ifmmode\stackrel{<}{_{\sim}}\else$\stackrel{<}{_{\sim}}$\fi}
\def\gapp{\ifmmode\stackrel{>}{_{\sim}}\else$\stackrel{<}{_{\sim}}$\fi}

\accepted{February 19 2020}

\shorttitle{The Sagittarius Stream in Gaia DR2}
\shortauthors{Ibata et al.}
\graphicspath{{./}{figures/}}

\begin{document}

\title{A panoramic landscape of the Sagittarius stream in Gaia DR2 \\
revealed with the {\tt STREAMFINDER} spyglass}

\correspondingauthor{Rodrigo Ibata}
\email{rodrigo.ibata@astro.unistra.fr}

\author[0000-0002-3292-9709]{Rodrigo Ibata}
\affiliation{Observatoire Astronomique, Universit\'e de Strasbourg, CNRS, 11, rue de l'Universit\'e, F-67000 Strasbourg, France}
\nocollaboration{1}

\author[0000-0001-8200-810X]{Michele Bellazzini}
\affiliation{INAF - Osservatorio di Astrofisica e Scienza dello Spazio, via Gobetti 93/3, I-40129 Bologna, Italy}
\nocollaboration{1}

\author[0000-0002-2468-5521]{Guillaume Thomas}
\affiliation{NRC Herzberg Astronomy and Astrophysics, 5071 West Saanich Road, Victoria, BC V9E 2E7, Canada}
\nocollaboration{1}

\author[0000-0002-8318-433X]{Khyati Malhan}
\affiliation{The Oskar Klein Centre, Department of Physics, Stockholm University, AlbaNova, SE-10691 Stockholm, Sweden}
\nocollaboration{1}

\author[0000-0002-1349-202X]{Nicolas Martin}
\affiliation{Observatoire Astronomique, Universit\'e de Strasbourg, CNRS, 11, rue de l'Universit\'e, F-67000 Strasbourg, France}
\affiliation{Max-Planck-Institut f\"ur Astronomie, K\"onigstuhl 17, D-69117, Heidelberg, Germany}
\nocollaboration{1}

\author[0000-0003-3180-9825]{Benoit Famaey}
\affiliation{Observatoire Astronomique, Universit\'e de Strasbourg, CNRS, 11, rue de l'Universit\'e, F-67000 Strasbourg, France}
\nocollaboration{1}

\author[0000-0001-8059-2840]{Arnaud Siebert}
\affiliation{Observatoire Astronomique, Universit\'e de Strasbourg, CNRS, 11, rue de l'Universit\'e, F-67000 Strasbourg, France}
\nocollaboration{1}

\begin{abstract}
We present the first full six-dimensional panoramic portrait of the Sagittarius stream, obtained by searching for wide stellar streams in the Gaia DR2 dataset with the {\tt STREAMFINDER} algorithm. We use the kinematic behavior of the sample to devise a selection of Gaia RR~Lyrae, providing excellent distance measurements along the stream. The proper motion data are complemented with radial velocities from public surveys. We find that the global morphological and kinematic properties of the Sagittarius stream are still reasonably well reproduced by the simple Law \& Majewski (2010) model (LM10), although the model overestimates the leading arm and trailing arm distances by up to $\sim 15$\%. The sample newly reveals the leading arm of the Sagittarius stream as it passes into very crowded regions of the Galactic disk towards the Galactic Anticenter direction. Fortuitously, this part of the stream is almost exactly at the diametrically opposite location from the Galactic Center to the progenitor, which should allow an assessment of the influence of dynamical friction and self-gravity in a way that is nearly independent of the underlying Galactic potential model.
\end{abstract}


\section{Introduction}
\label{sec:Introduction}

The Sagittarius dwarf galaxy \citep{1994Natur.370..194I} is one the major contributors to the stellar populations of the Galactic halo \citep{2002ApJ...569..245N,2006ApJ...642L.137B}. It is currently $\sim 19\kpc$ behind the Galactic bulge, and dissolving rapidly under the influence of the strong tides at that location. The tidally-disrupted stars that have been removed from the progenitor now form a vast, almost polar band, that wraps more than a full revolution around the sky \citep{2001ApJ...551..294I,2003ApJ...599.1082M}. It has long been appreciated that this system can inform us about the processes of minor mergers and satellite accretion, and the fundamental problem of the distribution of dark matter, both in the Milky Way and in its satellites.

Early simulations attempted to understand how such an apparently fragile system could survive to be seen at the present time, concluding that some dark matter component in the dwarf was probably necessary \citep{1998ApJ...500..575I}. The structure and kinematics of the stream indicated that the Galactic potential was roughly spheroidal, although different analyses concluded that the most likely shape was either spherical \citep{2001ApJ...551..294I}, slightly oblate \citep{2005ApJ...619..807L} or slightly prolate \citep{2004ApJ...610L..97H}, apparently dependent on the location of the tracers employed. 

A subsequent in-depth analysis \citep[][hereafter LM10]{2010ApJ...714..229L} of an all-sky survey of M-giant stars, found that a triaxial Galactic potential model could resolve these conflicts. The proposed model had the  surprising property of being significantly flattened along the Sun-Galactic center axis, which is difficult to reconcile with the dynamics of a stable disk configuration \citep{2013MNRAS.434.2971D}. Nevertheless, this model has held up remarkably well to subsequent observations of the Sagittarius system, and despite its limitations \citep{2016ASSL..420...31L} it has become the reference against which other models are held up \citep[see, e.g.,][]{2017A&A...603A..65T,2019MNRAS.483.4724F}.

Here we revisit this structure, using the superb new data from Second Data Release (DR2) of the Gaia mission \citep{2018A&A...616A...2L,2018A&A...616A...1G}. Our approach will be to use the {\tt STREAMFINDER} algorithm \citep{2018MNRAS.477.4063M,2018MNRAS.481.3442M} to identify stars that have a high likelihood of belonging to physically wide streams, such as that of the Sagittarius dwarf. Our aim is to provide the community with an effective means to select high-probability members of the stream from Gaia data.

\begin{figure*}
\begin{center}
\includegraphics[angle=0, viewport= 35 65 670 830, clip, width=16.5cm]{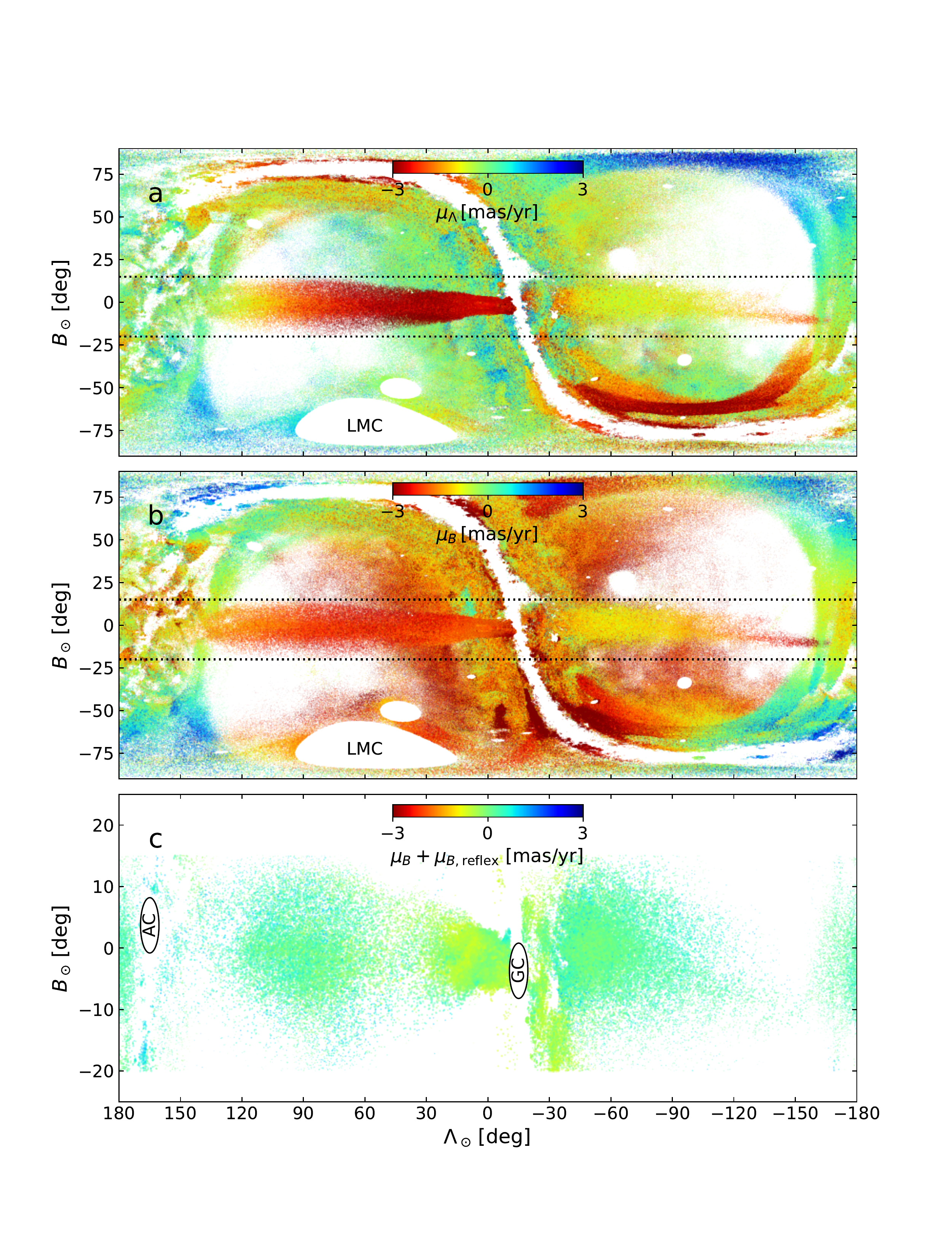}
\end{center}
\caption{Maps of {\tt STREAMFINDER} detections with stream significance $>15\sigma$ using a stream template of Gaussian width $0.5\kpc$ and a stellar populations template of age $12.5\Gyr$ and metallicity ${\rm [Fe/H]=-1.1}$. The $\Lambda_\odot, B_\odot$ Sagittarius coordinates shown are aligned such that $\Lambda_\odot$ points along the stream and $B_\odot$ is orthogonal to $\Lambda_\odot$. The proper motions in those directions, $\mu_\Lambda$ and $\mu_B$, are displayed on the top and middle panels. Panel c shows the $\mu_B$ proper motion corrected for Solar reflex motion for the final cleaned sample of 263,438 stars. Galactic satellites (e.g., the LMC) were masked-out in the input catalog. The positions of the Galactic Center and Galactic Anticenter are marked 'GC' and 'AC', respectively.} 
\label{fig:maps}
\end{figure*}

\section{{\tt STREAMFINDER} selection}
\label{sec:STREAMFINDER}

We re-analysed the Gaia DR2 dataset with the {\tt STREAMFINDER} algorithm in an almost identical way to the procedure described in \citet[][hereafter IMM19]{2019ApJ...872..152I}. As in IMM19, we only considered stars down to a limiting magnitude of $G_0=19.5$ (fainter sources were discarded to minimise spatial inhomogeneities in the maps). Magnitudes were corrected for extinction using the \citet{1998ApJ...500..525S} maps, adopting the re-calibration by \citet{2011ApJ...737..103S}, with $R_V=3.1$. The full sky was processed, although circular regions surrounding known satellites (but not the Sagittarius dwarf galaxy) were ignored. This masking of satellites is explained in detail in IMM19.

\begin{figure}
\begin{center}
\includegraphics[angle=0, viewport= 35 65 670 830, clip, width=\hsize]{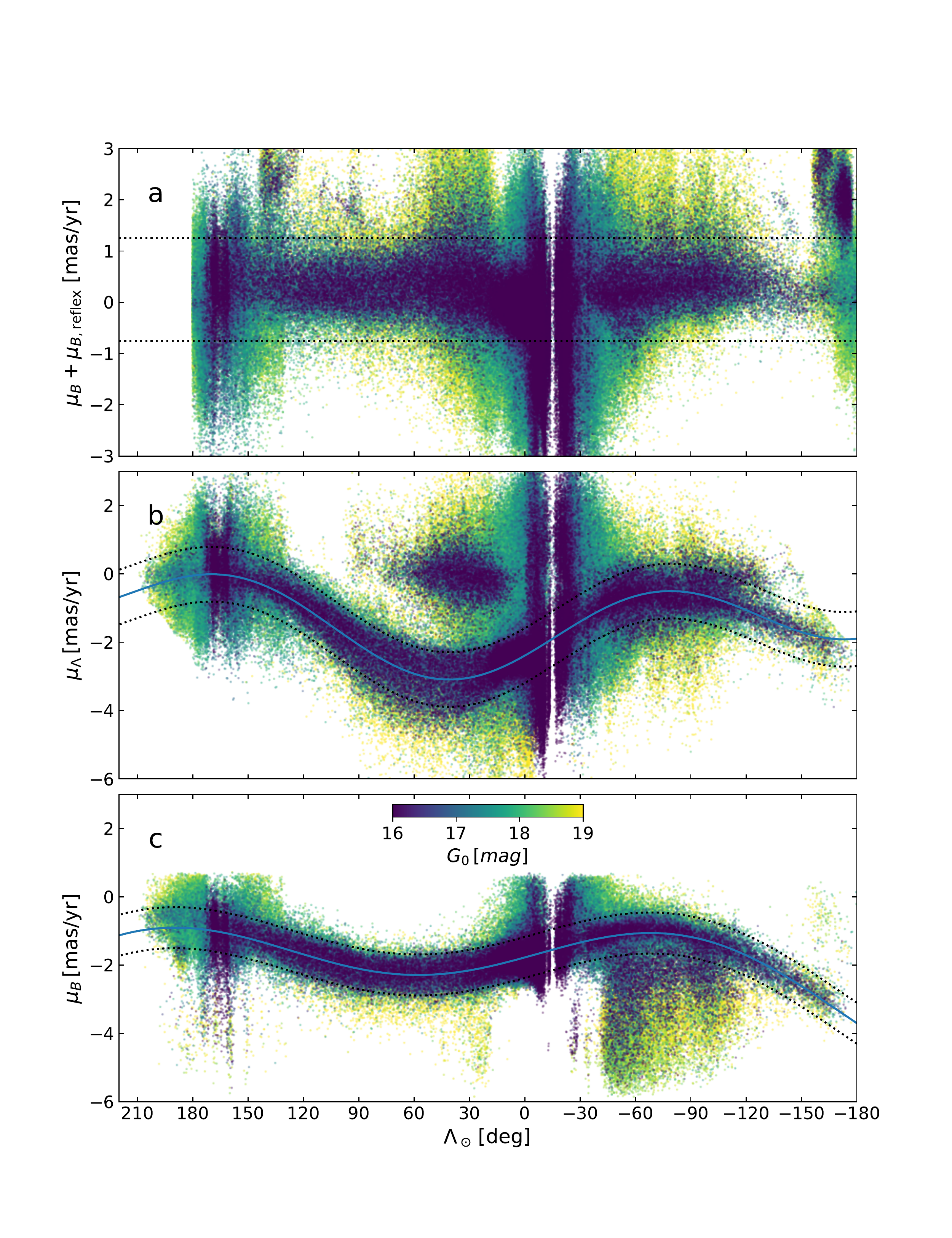}
\end{center}
\caption{Cleaning of the sample. After limiting the sources to $-20\deg<B<15\deg$, as shown in Figure~\ref{fig:maps}, we limit the sample to $-0.75< \mu_B+\mu_{B, \rm reflex}< 1.25\masyr$ (panel a). This constrains the stars to move along the orbit. b: The sample is trimmed further, taking $|\mu_\Lambda-\mu_{\Lambda, \rm fit}|<0.8\masyr$. c: Finally, we take $|\mu_B-\mu_{B, \rm fit}|<0.6\masyr$. The color of the points encode the $G_0$-band magnitude of the stars, and are shown here because the astrometric uncertainties are primarily a function of $G$.} 
\label{fig:selections}
\end{figure}

The {\tt STREAMFINDER} is effectively a friend-finding algorithm that considers each star in a dataset in turn, and searches for similar stars in a tube along all the possible orbits of the star under consideration (the orbits are integrated in the potential model \#1 of \citealt{1998MNRAS.294..429D}). The algorithm requires a stream template model as input. In the present work we adopted a stream width of (Gaussian) dispersion $0.5\kpc$, and allowed the algorithm to search for friends along a $20\deg$-long orbit. We ran the process with three different stellar populations templates from the PARSEC library \citep{2012MNRAS.427..127B} of age and metallicity $(T, {\rm [Fe/H]})$: $(8\Gyr,-1.4)$, $(12.5\Gyr,-1.1)$, $(12.5\Gyr,-1.7)$. Here we present the results using the $(12.5\Gyr,-1.1)$ model, which gave the best match to the RR~Lyrae distances calculated below. However, the samples derived from the three age-metallicity templates choices yield essentially identical proper motion profiles. The algorithm was only allowed to search for distance solutions in the Heliocentric range $d_\odot=[10,100]\kpc$. All other parameters were the same as in IMM19. These include using a Galactcentric distance of $R_\odot=8.122\pm0.031\kpc$ \citep{2018A&A...615L..15G}, and adopting a circular velocity of $v_c(R_\odot)=229.0\pm0.2\kms$  \citep{2019ApJ...871..120E}. Given that $v_c(R_\odot)+V_{\rm LSR, \, pec}+V_\odot=255.2\pm5.1 \kms$ \citep{2014ApJ...783..130R}, we take the sum of the $V$-component of the peculiar velocity of the Sun and $V$-component of the peculiar velocity of the Local Standard of Rest to be $V_\odot+V_{\rm LSR, \, pec}=26.2\kms$, while the $U$ and $W$ components of the Sun's peculiar velocity are taken from \citet{2010MNRAS.403.1829S}.

Figure~\ref{fig:maps} shows the resulting map of the stars in Gaia DR2 that exhibit stream-like behavior with significance $>15\sigma$. The $\Lambda_\odot, B_\odot$ coordinate system used is a version of the Heliocentric Sagittarius coordinates devised by \citet{2003ApJ...599.1082M}, although here we follow the choice of \citet{2012ApJ...750...80K} of inverting $B_\odot$ (so the maps are more easily compared to standard maps made in equatorial coordinates). The pole of the Great Circle is at $(\ell,b)=(273\degg8,-13\degg5)$, with zero-point of $\Lambda_\odot$ at the position of the globular cluster M54, commonly accepted to be the center of the system \citep{2008AJ....136.1147B}. The most recently-disrupted stars in the leading arm have negative values of $\Lambda_\odot$, and the dwarf galaxy is moving towards negative $\Lambda_\odot$. In panels (a) and (b) we display $\mu_\Lambda$ and $\mu_B$, respectively, which are the Gaia proper motion measurements rotated into these Sagittarius coordinates. The Sagittarius stream stands out as one of the most striking features in this all-sky map. It spans the entire sky and, as shown by  \citet{2006ApJ...642L.137B} and \citet{2012ApJ...750...80K}, it is bifurcated into two parallel arms over much of its length. Its varying width is a projection effect due to the large range of heliocentric distance it covers. Other known streams are present, and some potential new streams appear to have been detected, but we defer their analysis to a subsequent contribution. Visual inspection shows that the Sagittarius stream is present in the range $B=[-20\deg,15\deg]$ (between the dotted lines in Figure~\ref{fig:maps}), and henceforth we consider only those (755,343) stars that lie within this band.

While internally the {\tt STREAMFINDER} constructs associations between stars in a catalog, it proved to be impractical for computer memory reasons to store these links. Some post-processing is therefore required to disentangle the Sagittarius stream from other stream-like features. The adopted selection procedure is described in Figure~\ref{fig:selections}.

The stars in a stream will generally not have large motions in the direction perpendicular to the orbit, unless there are strong perturbers (see, e.g. \citealt{2019MNRAS.487.2685E}). For this reason, in Figure~\ref{fig:selections}a we conservatively take the broad selection in $-0.75\masyr < \mu_B+\mu_{B, \rm reflex} < 1.25\masyr$ of the proper perpendicular to the Sagittarius plane (corrected for the reflex motion $\mu_{B, \rm reflex}$ of the Sun in the direction of $B_\odot$). To compute $\mu_{B, \rm reflex}$, we assume the Galactic geometry and Solar motion described in IMM19, and use the distance to the stars that is estimated by the {\tt STREAMFINDER} software. The sample is clearly displaced with respect to the expected $\mu_B+\mu_{B, \rm reflex}=0\masyr$ line; the reason for this is unclear, but it may indicate that the stellar distances are underestimated (by $\sim 10$\%), or that the adopted model of the Solar motion is imprecise\footnote{\citet{2018ApJ...867L..20H} make use of this motion of the Sagittarius stream perpendicular to its plane to derive the Solar reflex motion, finding a value that is only $-2.2\kms$ lower than the value adopted in IMM19, and used here.}. This selection leaves 539,707 stars.

\begin{figure}
\begin{center}
\includegraphics[angle=0, viewport= 1 58 395 765, clip, width=\hsize]{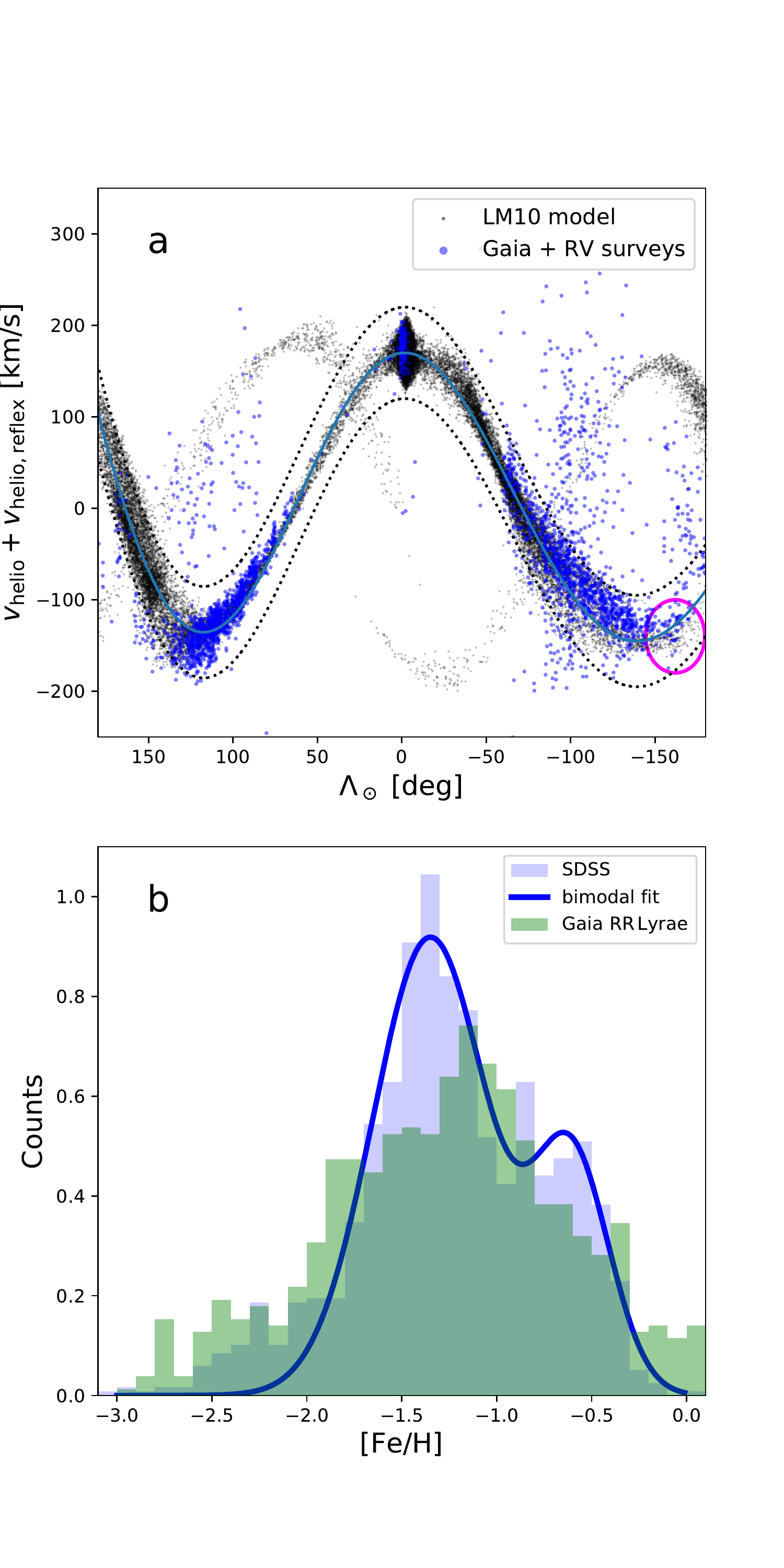}
\end{center}
\caption{a: Radial velocities of the cleaned {\tt STREAMFINDER} sample, as measured in public spectroscopic surveys. The behavior of the younger tidal arms in the ($<3\Gyr$) LM10 model are also shown, along with a sinusoidal fit to these particles (solid line). The magenta circle highlights the nearby leading arm. b: The metallicity distribution of the SDSS-Segue stars (blue) in the sample (and that also lie between the dotted lines in panel a) is compared to the Gaia RR~Lyrae sample (green). The blue line shows a bimodal Gaussian fit to the Segue sample.}
\label{fig:RV_and_met}
\end{figure}

In Figure~\ref{fig:selections}b, we show the subsequent selection on $\mu_\Lambda$. To model the sinusoidal behavior of the stream we fit a model to the brighter stars with $G_0<17.5$ using an iterative sigma-clipping procedure. The displayed model has the form:
\begin{equation}
\mu_{\Lambda, \rm fit}(\Lambda_\odot) = a_1 \sin(a_2 \Lambda_\odot + a_3) + a_4 +a_5 \Lambda_\odot + a_6 \Lambda_\odot^2
\end{equation}
and the best-fitting parameters (with $\Lambda_\odot$ in degrees) were found to be: $a_1=1.1842$, $a_2=-1.5639 \times \pi/180$, $a_3=-0.39917$, $a_4=-1.9307$, $a_5=-8.0606\times 10^{-4}$ and  $a_6=3.2441\times 10^{-5}$. The 331,795 stars with $|\mu_\Lambda - \mu_{\Lambda, \rm fit}| <0.8\masyr$ (a $2\sigma$ limit) were retained. 

We also make a selection on $\mu_B$, as shown in Figure ~\ref{fig:selections}c. The fitted function, $\mu_{B, \rm fit}(\Lambda_\odot)$ has the same functional form as $\mu_{\Lambda, \rm fit}(\Lambda_\odot)$, but with parameters $a_1=-1.2360$, $a_2=1.0910 \times \pi/180$, $a_3=0.36330$, $a_4=-1.3412$, $a_5=7.3022\times 10^{-3}$ and  $a_6=-4.3315\times 10^{-5}$. By selecting $|\mu_B - \mu_{B, \rm fit}| <0.6\masyr$ (again a $2\sigma$ limit), we obtain a final sample of 263,438 candidate stream stars. The spatial distribution of these sources is displayed in Figure~\ref{fig:maps}c, and they are listed in Table~\ref{tab:data}.

We cross-matched these sources with public spectroscopic surveys, and found 2984 matches (212 in APOGEE, \citealt{2017AJ....154...94M}; 35 in the Gaia Radial Velocity Spectrometer sample; 1236 in LAMOST, \citealt{2012RAA....12.1197C}; 1 in RAVE, \citealt{2017AJ....153...75K}; and 1500 in SDSS-Segue, \citealt{2009AJ....137.4377Y}). The radial velocities of these stars are shown in Figure~\ref{fig:RV_and_met}a along with the LM10 simulation. The improvement in terms of contamination in this {\tt STREAMFINDER} sample over large pre-Gaia surveys can be appreciated by comparing Figure~\ref{fig:RV_and_met}a to the SDSS study by \citet{2017MNRAS.464..794G} (their Figure~1). The sub-sample with radial velocities can be used as a control sample to estimate the contamination fraction. To this end, we fit a sinusoid to the young ($<3\Gyr$) arms of the Sagittarius stream in the LM10 model (blue line), and assume that stars beyond $50\kms$ (dotted lines) of this fit are Galactic field star contaminants. Given the velocity dispersion of metal-poor stars in the stream ($13\kms$, \citealt{2017MNRAS.464..794G}), this corresponds to a $\sim 4\sigma$ limit, that is wide enough to allow for some model mismatch. The resulting contamination fraction is 18\%. Note however, that this is a global value, averaged over the very complex footprint and complex target selection functions of the public radial velocity surveys listed above. Breaking down this test sample by magnitude, we find a contamination fraction of 14\% for $G_0<17$~mag; of 19\% for $G_0$ in the range $[17,18]$~mag, and of $30\%$ for $G_0>18$~mag. Clearly, the contamination fraction will be dependent on the density of the contaminating populations, and so will be highest at low Galactic latitude. In Figure~\ref{fig:maps}c, the off-track population with $B_\odot<-10\deg$ and $\Lambda_\odot$ in the range $[-40\deg,-20\deg]$ (and which straddles the Galactic plane behind the bulge) looks suspiciously like such contamination.

It is very difficult to predict the effect that the false positives may have on subsequent kinematic analyses, but given that the contamination fraction is relatively small the effect may be small also. Selecting stars closer to the fitted proper motion track helps to reduce the contamination. Taking $|\mu_\Lambda-\mu_{\Lambda, \rm fit}|<0.4\masyr$ and $|\mu_B-\mu_{B, \rm fit}|<0.3\masyr$ (i.e. tightening the previous constraints by a factor of 2), yields a sample of 138,165 stars with a contamination fraction of 11\%, and estimated in the same way as above). 

Considering the sub-sample of stars possessing SDSS radial velocity and metallicity measurements, we find $\langle {\rm [Fe/H]} \rangle =-1.24$~dex ($-1.40$~dex) and $\sigma_{\rm [Fe/H]}=0.52$~dex ($0.60$~dex) for the velocity-confirmed members (non-members). The similarity of the metallicity distributions of the stream and the contaminants means that metallicity can only be weakly correlated with the contamination probability. Furthermore, the correlation between  $B_\odot$ and ${\rm [Fe/H]}$ is very low, with a Spearman's rank coefficient of $\rho=0.004$, and $\rho=-0.024$ for the correlation between $|B_\odot|$ and ${\rm [Fe/H]}$. This confirms that the populations of different metallicities are not appreciably displaced perpendicular to the stream track (and shows again that metallicity cannot be a primary driver of contamination probability).

One might be concerned that the stellar populations template used in the {\tt STREAMFINDER} could introduce a strong bias against stars of different metallicity. This is not the case, as we show in Figure~\ref{fig:RV_and_met}b, where we display the metallicity distribution of the sample with SDSS-Segue metallicities and that are confirmed velocity members (blue). Fitting a bimodal Gaussian to these data (blue line) yields means of 
${\rm [Fe/H]}=-1.35$ and ${\rm [Fe/H]}=-0.61$ with metallicity dispersions of $0.30$~dex and $0.20$~dex, respectively. These values are extremely close to the trailing arm fit by \citet{2017MNRAS.464..794G}: ${\rm [Fe/H]}=-1.33$ and ${\rm [Fe/H]}=-0.74$ with dispersions $0.27$~dex and $0.18$~dex, respectively. Thus the {\tt STREAMFINDER} sample does not have an obvious metallicity bias.

The Gaia DR2 catalog is known to have a small $0.029$~mas parallax bias \citep{2018A&A...616A...2L}. Assuming that the correlation matrix of the astrometric solution is valid in the context of this small parallax bias, we can use the Gaia {\tt parallax\_pmra\_corr} and {\tt parallax\_pmdec\_corr} terms to derive the resulting proper motion bias. The resulting mean bias values for the present sample are $-0.005\masyr$ (rms scatter $0.007\masyr$) and $0.002\masyr$ (rms scatter $0.006\masyr$) for the bias in $\mu_\alpha$ and $\mu_\delta$, respectively. If there are any applications of this dataset that need a mean accuracy beyond this level, they will need to update the proper motion values in Table~\ref{tab:data} using the Early Data Release 3 catalog (expected for late 2020).

In a recent analysis of the GD-1 stellar stream, we showed that a sample derived with the {\tt STREAMFINDER} software had a statistically-identical density profile to samples defined in a more traditional way by sigma-clipping, followed by background subtraction \citep{2020arXiv200201488I}. Thus, at least for mono-metallicity populations, the algorithm does not produce samples with peculiar completeness properties. However, in the present work, we cleaned the initial {\tt STREAMFINDER} sample with the three proper motion filters depicted in Figure~\ref{fig:selections} in order to better isolate the Sagittarius stream stars and reduce the number of false positives. Unfortunately, the proper motion uncertainties of fainter stars will cause genuine members to drop out of the selection windows (we note that the most stringent cut of $0.6\masyr$ on $|\mu_B-\mu_{B, \rm fit}|$ corresponds to the typical proper motion uncertainty at $G_0 \sim 19$~mag). This will lead to an increasing incompleteness of the faint stars. Studies that require sample completeness will need to correct for this loss of members. We estimate the incompleteness caused by these three proper motion filters by applying them to a version of the LM10 model where the N-body particles are assigned a $G$-band magnitude drawn from the PARSEC model with $(T, {\rm [Fe/H]})=(12.5\Gyr,-1.1)$. Only the particles within $180\deg$ of the progenitor are considered (i.e. we neglect older wraps). Proper motion uncertainties are assigned as a function of $G$ using the median values listed in \citet{2018A&A...616A...2L}, and the model proper motions are degraded accordingly. We thereby find that the global incompleteness caused by the three proper motion filters is $<5$\% to $G=17.5$~mag, but degrades to $7$\% for $G=[17.5,18.5]$~mag, and to $40$\% for $G=[18.5,19.5]$~mag.

\section{Sagittarius stream RR~Lyrae stars in Gaia}
\label{sec:RRLyrae}

The spatial and proper motion selection procedure described above also provides a means to construct a cleaned catalog of Sagittarius RR~Lyrae stars, which can serve as distance anchors to the stream. For this we used the RR~Lyrae variables identified in the {\tt gaiadr2.vari\_rrlyrae} catalog \citep{2019A&A...622A..60C}, that is part of Gaia DR2. The catalog includes 140,784 RR~Lyrae and provides a metallicity estimate from Fourier parameters of the light curves (see, e.g., \citealt{2013ApJ...773..181N}) for 64,932 of them. From this source we selected the subset of 135,825 stars having full 5-parameter astrometric solution. Interstellar extinction was corrected for in the same manner as described above for the main Gaia catalog. After applying the selection on $B_\odot$, as well as the proper motion selections presented in Figure~\ref{fig:selections}, we obtain a cleaned sample of (exactly) 3,500 Sagittarius RR~Lyrae stars. 

\begin{figure*}
\begin{center}
\includegraphics[angle=0, viewport= 65 110 820 1170, clip, width=14cm]{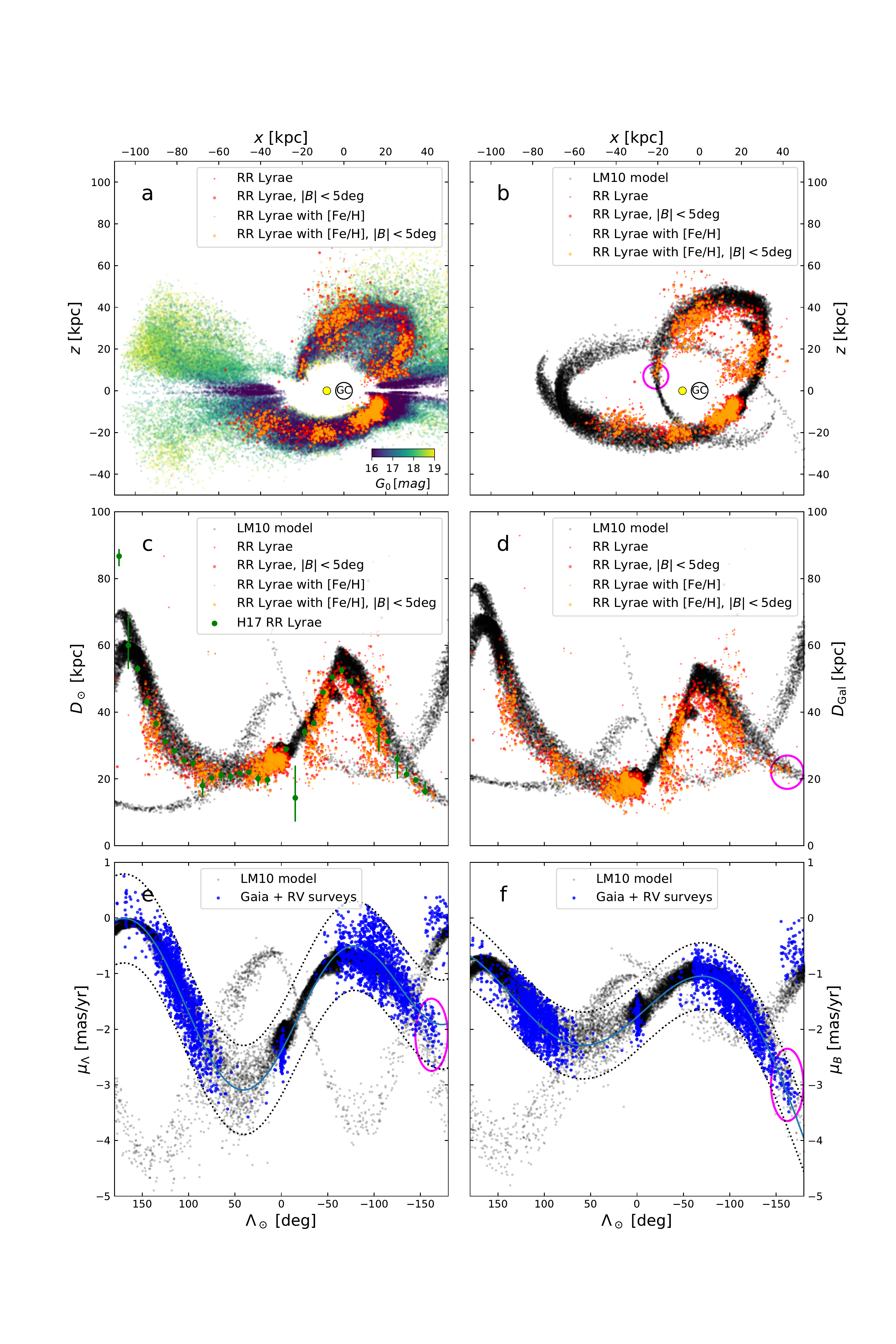}
\end{center}
\caption{a: The $x$-$z$ plane positions of the cleaned {\tt STREAMFINDER} sample (colored by their $G_0$ magnitude) are compared to the Gaia RR~Lyrae stars. Note the good match both to the RR~Lyrae stars with metallicity measurements (orange) and without (red). Larger dots mark the stars with $|B|<5\deg$ that better delineate the stream. b: The position of the RR~Lyrae are compared to the LM10 model, showing some significant systematic discrepancies both in the leading and trailing arms. The magenta circle highlights the nearby portion of the leading stream seen towards the Galactic Anticenter. (The position of the Sun is marked with a yellow circle, and the Galactic Center is encircled in black). c: Profile in Heliocentric distance. d: Profile in Galactocentric distance. The disagreement in distance with the LM10 model can be seen more clearly here. Note, however, that the model agrees well at the nearby section of the leading arm (highlighted in magenta). Comparison of the proper motion profiles in $\mu_\Lambda$ (e) and $\mu_B$ (f) between the spectroscopic {\tt STREAMFINDER} sample and the LM10 model. The sinusoidal selection functions from Figures~\ref{fig:selections}b and \ref{fig:selections}c have been overlaid in panels e and f, respectively.}
\label{fig:combined}
\end{figure*}

For the subset of stars for which metallicity estimates are available in the catalog (1,474 stars), we calculate the distance from the $M_G$ -- ${\rm [Fe/H]}$ relation by \citet{2018MNRAS.481.1195M}. The metallicity distribution of this RR~Lyrae sample is displayed in Figure~\ref{fig:RV_and_met}b (green); a large metallicity spread is present, but this is also seen in the SDSS-Segue {\tt STREAMFINDER} sample (blue). The mean metallicity of the RR~Lyrae is ${\rm [Fe/H]=-1.3}$, corresponding to $M_G=+0.69$, which we adopted for all the RR~Lyrae in the sample lacking metallicity. To provide a quantitative idea of the size of the systematic error possibly associated with this choice, the adoption of  $M_G=0.64$, following \citet{2019MNRAS.482.3868I}, would lead to a distance scale larger than ours by $2.5$\%, a negligible amount in the present context. In Figure~\ref{fig:combined}c we show (in green) the distances to the stream derived from RR~Lyrae identified in Pan-STARRS \citep[][hereafter H17]{2017ApJ...850...96H}. The slight differences as a function of position may be due to the fact that the Gaia RR~Lyrae sample is much less contaminated and that the metallicity correction applied here --- but which H17 could not implement due to a lack of metallicity information --- improves the distances.

The $x$--$z$ plane positions of these stars are compared to the values calculated by the {\tt STREAMFINDER} in Figure~\ref{fig:combined}a; the good match shows that the {\tt STREAMFINDER} provides useful estimates of the distance to the Sagittarius stream with the adopted stellar population template.

\section{Discussion and Conclusions}
\label{sec:Discussion}

We now compare these new data to the N-body simulation by LM10, which has proved over the years to be an extremely useful model. Here we consider only those particles that were disrupted and became gravitationally unbound from the progenitor $3\Gyr$ ago, or less.  Figures~\ref{fig:combined}b,c,d show that the distances to the particles in the LM10 model are substantially overestimated, by up to $\sim 15$\% along large portions of the leading arm. In contrast, the model follows closely the proper motion behavior of the stream (Figures~\ref{fig:combined}e,f)\footnote{We note that revising down $v_c(R_\odot)+V_{\rm LSR, \, pec}+V_\odot$ to $247\kms$ \citep{2019ApJ...885..131R} changes the position of the model particles on average by $-0.02\masyr$ in $\mu_\Lambda$ and $0.09\masyr$ in $\mu_B$.}, although systematic offsets (of up to $\sim 0.2\masyr$) are present in both arms (and tend to be particularly pronounced in regions where the distances are overestimated). The match in radial velocity is also good (Figure~\ref{fig:RV_and_met}a), although some discrepancies are also apparent, for instance in the trailing arm at $\Lambda_\odot \sim 130\deg$, where the model over-predicts the radial velocity by $\sim 50\kms$.

Inspection of Figure~\ref{fig:combined}b suggests that the distance discrepancy with the LM10 model starts at the very base of the leading arm, hinting that the L1 Lagrange point may not be sufficiently close to the Milky Way center. Note that LM10 take the distance from the Sun to the Sagittarius dwarf to be $D=28\kpc$, which is the largest value used in the literature, for instance it is $\sim 6$\% larger than the distance to M54 quoted by \citet{2010arXiv1012.3224H}. We expect that a better fit to the distances along the leading arm may result from decreasing $D$ and increasing the mass of the Milky Way model (note also that the LM10 simulations adopted a model with a speed of the local standard of rest of $v_{\rm LSR}=220\kms$, which is substantially lower than currently-preferred values). A comprehensive suite of numerical simulations is now needed to properly explore the parameter space of Milky Way potential models as well as models for the Sagittarius dwarf galaxy itself. This is, however, beyond the scope of the present letter.

We will instead now focus on an interesting feature of the stream, highlighted in Figures~\ref{fig:RV_and_met}--\ref{fig:combined} with a magenta circle. This part of the Sagittarius stream corresponds to the location where the leading arm plunges down into the Galactic disk, in the direction of the Galactic Anticenter, and where it is closest to us. As can be seen in Figure~\ref{fig:combined}b, the LM10 model accurately predicted the location of this feature, $\sim 21\kpc$ away from the Galactic center. As the stars speed up on their long trajectory falling almost vertically onto the disk, the conservation of phase space density (encapsulated in Liouville's theorem) causes a ``pinching'' of the stream in configuration space, as is observed.

In the coordinate system of Figures~\ref{fig:combined}a,b, the feature is located at $\vec{x}_1 \sim (-20,-5,6)\kpc$, approximately at the diametrically opposite location to the Sagittarius dwarf $\vec{x}_0 = (17.5,2.5,-6.4)\kpc$ (taking values for M54 from \citealt{2010arXiv1012.3224H}). Assuming that the Galactic potential has the symmetry $\Phi(\vec{x})=\Phi(-\vec{x})$ (which is the case in the LM10 potential or indeed in any fixed triaxial potential as long as one of the principal axes is perpendicular to the Galactic plane), the difference in total velocity between the progenitor at $\vec{x}$ and its stream at $-\vec{x}$ should only be due to the effect of dynamical friction of the remnant and self-gravity in the stream. The proximity of $\vec{x}_1$ to $\vec{x}_0$ in the potential can be appreciated by noting that in the LM10 potential model, if a test particle moves in a ballistic orbit from $\vec{x}_0$ starting with the velocity magnitude of the Sagittarius dwarf ($321\kms$), its velocity magnitude decreases by 5.8\% when reaching $\vec{x}_1$ (the same value of 5.8\% is obtained with the potential model \#1 of \citealt{1998MNRAS.294..429D}).

We suspect that it will be possible to use this approximate property of the nearby stream to constrain the total mass of the Sagittarius dwarf over the period of time since those stars were detached from the progenitor ($\sim 3\Gyr$ in the LM10 model). The reason this is promising is that it should allow us to isolate energy differences due to dynamical friction and self-gravity from energy differences due to position in the potential. This would simplify greatly the parameter space of N-body models that need to be surveyed to reproduce the Sagittarius system.

Finally, we cannot help but note that LM10 constructed what is still the best model of the Sagittarius stream, following the observed properties of the structure in terms of position, distance and kinematics (Figures~\ref{fig:RV_and_met}--\ref{fig:combined}). This includes predictions for portions of the stream that were not known in 2010, as well as for proper motions and distances that have improved enormously in the intervening years. The LM10 simulations did not account for dynamical friction (as they did not include a live halo), but given that they used a progenitor model of initial mass $6.4\times 10^8\msun$, dynamical friction could be neglected. In contrast, modern abundance-matching arguments assign the Sagittarius dwarf galaxy to the third most massive sub-halo in the Milky Way system, leading to mass estimates (at infall) of $5.7\times 10^{10}\msun$ \citep{2019MNRAS.487.5799R}. Detailed live simulations have shown that such masses at infall are indeed required in models where the Sagittarius galaxy excites, flares, bends and corrugates the Galactic disk \citep[e.g.,][]{2018MNRAS.481..286L} to reproduce the locations and motions of feathers and arc-like overdensities in the outer Milky Way disk. The fact that the LM10 model, two orders of magnitude lower in mass, matches observations as well as it does, means that the combination of the modelled potential and the modelled self-gravity somehow mimic the combination of the real potential, the real self-gravity, the real dynamical friction, and the real perturbations (in particular, from the Large Magellanic Cloud). In future work it will be interesting to verify quantitatively  that massive models can also reproduce the observed large-scale six-dimensional phase-space structure of the Sagittarius stream.

As the present letter was being reviewed, \citet{2020arXiv200110012A} published a sample of Sagittarius stream stars derived from Gaia DR2 data. Their identification technique is very different from that presented here, but our analyses appear to give globally consistent results.

\begin{table*}
\caption{The first 10 rows of the {\tt STREAMFINDER} sample of 263,438 stars in the Sagittarius Stream.}
\label{tab:data}
\begin{tabular}{ccccccccccc}
\hline
\hline
$\alpha$ & $\delta$ & $\mu_\alpha$ & $\mu_\delta$ & $G_0$ & $(G_{BP}-G_{RP})_0$ & $d_{SF}$ & $\Lambda_\odot$ & $B_\odot$ & $\mu_\Lambda$ & $\mu_B$ \\
$\deg$ & $\deg$ & $\masyr$ & $\masyr$ & mag & mag & $\kpc$ & $\deg$ & $\deg$ & $\masyr$ & $\masyr$ \\
    0.001305 &   -24.216246 &    -1.605 &    -3.730 &    18.491 &     0.983 &    24.602 &    66.770 &    -5.626 &    -3.034 &    -2.699 \\
    0.001397 &   -25.892900 &    -1.445 &    -3.001 &    17.708 &     1.014 &    19.603 &    66.054 &    -7.144 &    -2.583 &    -2.102 \\
    0.004864 &    -4.348376 &    -1.480 &    -3.457 &    18.862 &     0.820 &    16.639 &    75.248 &    12.366 &    -2.827 &    -2.480 \\
    0.005143 &   -31.402405 &    -1.721 &    -3.100 &    17.031 &     1.010 &    18.122 &    63.666 &   -12.125 &    -2.890 &    -2.055 \\
    0.006523 &   -24.277519 &    -1.460 &    -3.738 &    18.562 &     0.929 &    17.315 &    66.748 &    -5.683 &    -2.906 &    -2.767 \\
    0.011688 &   -22.100415 &    -1.761 &    -3.350 &    18.206 &     1.038 &    23.118 &    67.676 &    -3.712 &    -3.011 &    -2.293 \\
    0.018644 &   -17.789699 &    -1.423 &    -2.943 &    16.138 &     1.181 &    19.071 &    69.502 &     0.193 &    -2.531 &    -2.069 \\
    0.019391 &   -23.791326 &    -1.660 &    -3.009 &    19.056 &     0.900 &    19.368 &    66.965 &    -5.248 &    -2.778 &    -2.024 \\
    0.020223 &   -20.515382 &    -1.588 &    -2.617 &    18.144 &     1.054 &    26.416 &    68.354 &    -2.279 &    -2.544 &    -1.703 \\
    0.021636 &   -27.217454 &    -1.312 &    -2.726 &    17.524 &     1.110 &    20.564 &    65.502 &    -8.350 &    -2.348 &    -1.907 \\
\hline
\hline
\end{tabular}
\tablecomments{Columns 1--6 list the Gaia equatorial coordinates $\alpha$ and $\delta$, proper motions $\mu_\alpha (* \cos(\delta))$, $\mu_\delta$, magnitude $G_0$, and color $(G_{BP}-G_{RP})_0$. The extinction correction is explained in the text. Column 7 provides the distance estimate $d_{SF}$ provided by the {\tt STREAMFINDER}. Finally, columns 8--11 give the same information as columns 1--4, but rotated into the Sagittarius coordinate system.}
\end{table*}

\acknowledgments

RI, NM, BF and AS acknowledge funding from the Agence Nationale de la Recherche (ANR project ANR-18-CE31-0006, ANR-18-CE31-0017 and ANR-19-CE31-0017), from CNRS/INSU through the Programme National Galaxies et Cosmologie, and from the European Research Council (ERC) under the European Unions Horizon 2020 research and innovation programme (grant agreement No. 834148).

MB acknowledges the financial support to this research by INAF, through the Mainstream Grant 1.05.01.86.22 assigned to the project "Chemo-dynamics of globular clusters: the Gaia revolution” (P.I. E. Pancino).

This work has made use of data from the European Space Agency (ESA) mission {\it Gaia} (\url{https://www.cosmos.esa.int/gaia}), processed by the {\it Gaia} Data Processing and Analysis Consortium (DPAC, \url{https://www.cosmos.esa.int/web/gaia/dpac/consortium}). Funding for the DPAC has been provided by national institutions, in particular the institutions participating in the {\it Gaia} Multilateral Agreement. 

Funding for SDSS-III has been provided by the Alfred P. Sloan Foundation, the Participating Institutions, the National Science Foundation, and the U.S. Department of Energy Office of Science. The SDSS-III web site is http://www.sdss3.org/.

SDSS-III is managed by the Astrophysical Research Consortium for the Participating Institutions of the SDSS-III Collaboration including the University of Arizona, the Brazilian Participation Group, Brookhaven National Laboratory, Carnegie Mellon University, University of Florida, the French Participation Group, the German Participation Group, Harvard University, the Instituto de Astrofisica de Canarias, the Michigan State/Notre Dame/JINA Participation Group, Johns Hopkins University, Lawrence Berkeley National Laboratory, Max Planck Institute for Astrophysics, Max Planck Institute for Extraterrestrial Physics, New Mexico State University, New York University, Ohio State University, Pennsylvania State University, University of Portsmouth, Princeton University, the Spanish Participation Group, University of Tokyo, University of Utah, Vanderbilt University, University of Virginia, University of Washington, and Yale University.

Guoshoujing Telescope (the Large Sky Area Multi-Object Fiber Spectroscopic Telescope LAMOST) is a National Major Scientific Project built by the Chinese Academy of Sciences. Funding for the project has been provided by the National Development and Reform Commission. LAMOST is operated and managed by the National Astronomical Observatories, Chinese Academy of Sciences.

\vfill\eject

\bibliography{Gaia_Sgr}
\bibliographystyle{aasjournal}

\end{document}